%% file: main.tex
\documentclass[9pt, conference]{IEEEtran}

\usepackage{graphicx} 
\usepackage[utf8]{inputenc}
\usepackage[T1]{fontenc}
\usepackage{microtype}
\usepackage{xargs}
\usepackage{lipsum}
\usepackage{xparse}
\usepackage{xifthen, xstring}
\usepackage{xspace}
\usepackage{marginnote}
\usepackage{etoolbox}
\usepackage{soul}
\usepackage{cite}
\usepackage{amsmath,amsfonts}
\usepackage{algorithmic}
\usepackage{graphicx}
\usepackage{textcomp}
\usepackage{hhline}
\usepackage{multirow}
\usepackage{booktabs}
\usepackage{subcaption}
\usepackage{tabularray}
\usepackage{pifont}
\usepackage[dvipsnames]{xcolor}
\usepackage{float}
\usepackage{colortbl}
\usepackage{tikz}
\usepackage{threeparttable}
\usepackage{tablefootnote}
\usepackage{adjustbox}
\usepackage{amssymb}
\usepackage{pifont} 
\usepackage{hyperref}

\usepackage{xcolor}
\definecolor{darkgreen}{RGB}{0,100,0}

\usepackage{tabularx}
\usepackage{subcaption}
\usepackage{multicol}
\usepackage{tikz}
\usepackage{stfloats}

\title{An Open-Source HW-SW Co-Development Framework Enabling Efficient Multi-Accelerator Systems}
\date{August 2025}

\IEEEoverridecommandlockouts 
\IEEEpubid{\makebox[\columnwidth]{ 979-8-3315-2710-5/25/\$31.00 ©2025 IEEE\hfill{ \hspace{\columnsep} \makebox[\columnwidth]{ }}}}

\begin{document}

\author{
    \IEEEauthorblockN{Ryan Albert Antonio$^*$, Joren Dumoulin$^*$, Xiaoling Yi, Josse Van Delm, Yunhao Deng, Guilherme Paim, Marian Verhelst}
    \IEEEauthorblockA{
        MICAS-ESAT, KU Leuven, Belgium \\
        ryan.antonio@esat.kuleuven.be, joren.dumoulin@esat.kuleuven.be
    }
}

\maketitle

\def\thefootnote{*}\footnotetext{Both authors contributed equally to this research}\def\thefootnote{\arabic{footnote}}

\input{0-abstract}

\input{1-introduction}
\input{2-background}

\input{2.5-snax}
\input{3-hardware}
\input{4-software}
\input{5-evaluation}

\input{6-conclusion}

\section*{Acknowledgment}

This project has been partly funded by the European Research Council (ERC) under grant agreement No. 101088865, the European Union’s Horizon 2020 program (CONVOLVE) under grant agreement No. 101070374, the Flanders AI Research Program, and KU Leuven.

\bibliographystyle{IEEEtran}
\bibliography{refs}


\end{document}

%% file: 0-abstract.tex
\begin{abstract}

Heterogeneous accelerator-centric compute clusters are emerging as efficient solutions for diverse AI workloads.
However, current integration strategies often compromise data movement efficiency and encounter compatibility issues in hardware and software.
This prevents a unified approach that balances performance and ease of use.
To this end, we present SNAX, an open-source integrated HW-SW framework enabling efficient multi-accelerator platforms through a novel hybrid-coupling scheme, consisting of loosely coupled asynchronous control and tightly coupled data access.
SNAX brings reusable hardware modules designed to enhance compute accelerator utilization, and its customizable MLIR-based compiler to automate key system management tasks, jointly enabling rapid development and deployment of customized multi-accelerator compute clusters. 
Through extensive experimentation, we demonstrate SNAX’s efficiency and flexibility in a low-power heterogeneous SoC. Accelerators can be easily integrated and programmed to achieve $> 10 \times$ improvement in neural network performance compared to other accelerator systems while maintaining accelerator utilization of $> 90\%$ in full system operation.

\end{abstract}

%% file: 1-introduction.tex
\section{Introduction}
\label{sec:introduction}

Specialization is essential for boosting computational efficiency in embedded devices running machine learning algorithms \cite{overview_cpu_history}.
The rapid growth of AI applications demands scalable and efficient systems to handle complex workloads. Accelerators improve efficiency through task-specific hardware \cite{overview_hpc_hardware_acc}, while multi-core systems enable scalability by distributing workloads across processing units \cite{overview_xcentric}.
However, the diverse and specialized nature of contemporary workloads requires heterogeneous systems that integrate these approaches and combine the benefits of task-specific acceleration with scalable multi-core processing to achieve optimal system performance.

As hardware and software systems become increasingly complex, integrating and managing multiple accelerators into scalable heterogeneous systems poses major difficulties \cite{esp_agile_dev}.
From the control perspective, it involves dispatching workloads to accelerators through varying interfacing methods, scheduling tasks, and synchronizing multiple cores.
From a data perspective, on the other hand, providing high-bandwidth connections, managing transfers, reducing congestion, and ensuring data coherence are critical to performance but risk bottlenecking accelerators.
Leveraging the hardware fully is complex and time-intensive, often due to the custom nature of these architectures.

Existing programmable multi-accelerator systems, such as ESP \cite{esp_agile_dev} and X-HEEP \cite{xheep_epfl} mainly emphasize the integration of hardware accelerators within heterogeneous systems.
However, these systems often adopt a hardware-centric approach, placing the burden of system management on software developers through low-level directives. By prioritizing hardware considerations, this approach overlooks critical software requirements during system design, complicating the development of compilers for heterogeneous architectures. As a result, the lack of a cohesive hardware-software integration strategy hampers accelerator utilization and reduces overall system efficiency. 

High-level compilers could ease these challenges through automated deployment, but such tools currently exist only for general-purpose CPUs, GPUs, or very specific accelerator architectures \cite{chipyard_gemmini, aha_stanford_project, pulp_dory}. 
There is a pressing need for a design framework that bridges the gap between efficient accelerator integration and ease of deployment, addressing hardware integration and software programmability. 

\begin{figure}[t]
    \centering
    \includegraphics[width=0.95\columnwidth]{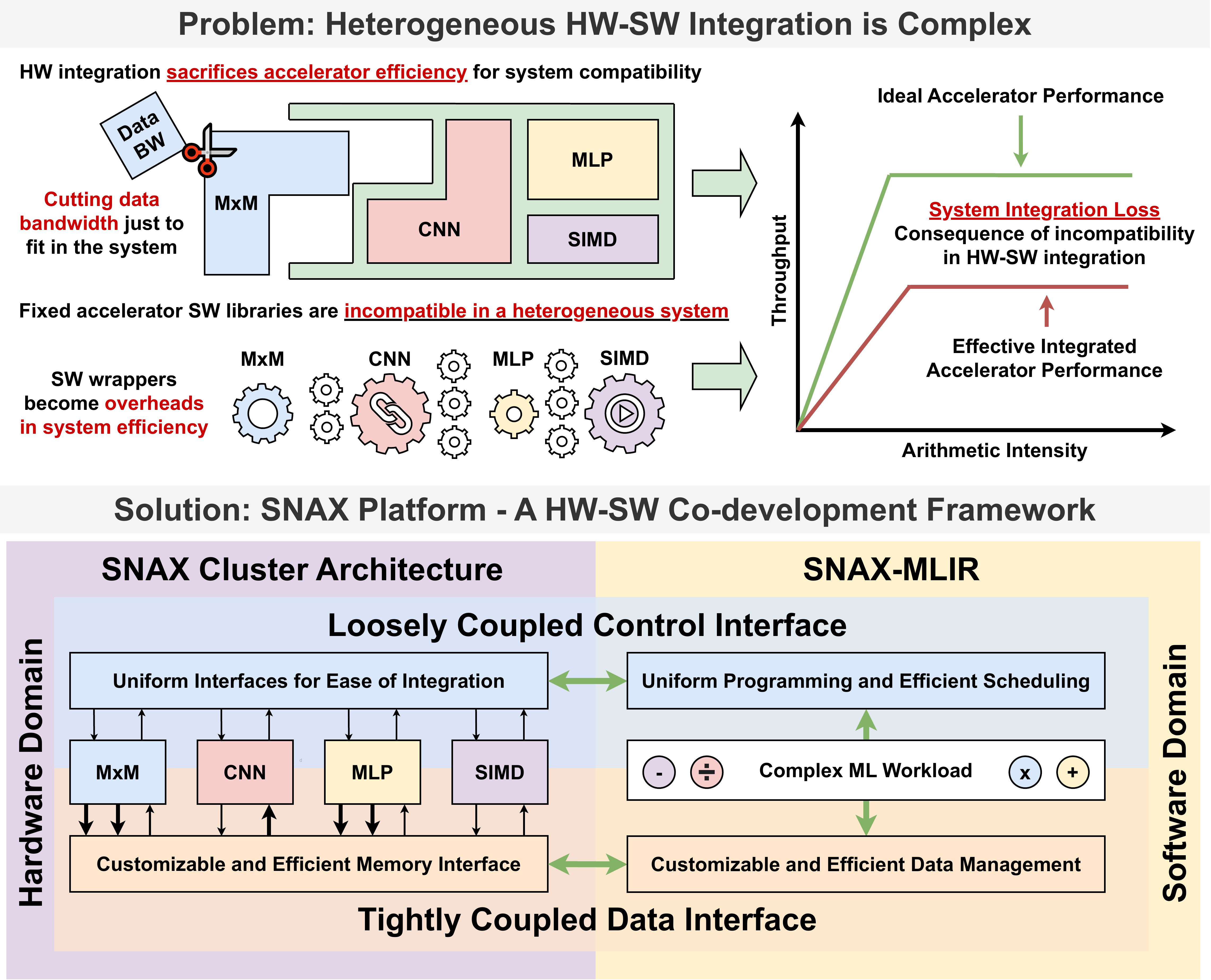}
    \caption{Challenges in HW-SW integration and the envisioned HW-SW co-development framework}
    \label{fig:SNAX_vision}
\end{figure}

Our key insight is that a well-defined, open-source development template acting as an abstraction layer between hardware and software can effectively address these challenges but is currently missing for heterogeneous accelerator architectures.
Similar to how the RISC-V architecture provides an abstraction for general-purpose processors, such a model for (multi-)accelerators would empower compiler engineers with precise knowledge of hardware behavior, enabling them to optimize software effectively. Simultaneously, within this hardware model, hardware engineers can optimize and extend the system for maximum efficiency while leveraging the existing framework to integrate other accelerators smoothly. 

To realize this vision, we propose \textbf{SNAX} (Figure \ref{fig:SNAX_vision}), a multi-accelerator compute cluster template that enables rapid integration of new accelerators while preserving accelerator efficiency up to the system level.  
Through a hybrid data/control coupling approach—combining a loosely coupled control interface with a tightly coupled data interface—SNAX serves as an effective framework for a HW-SW co-developed system.

Building on the SNAX template, we developed both a HW architecture and an MLIR-based compiler toolchain to bring this concept to fruition: 

\begin{itemize} 
    \item \textbf{SNAX Cluster}: An open-source multi-accelerator compute cluster that can easily integrate custom accelerators, has a highly customizable architecture with supporting modules to enhance.  
    \item \textbf{SNAX-MLIR}: An MLIR-based compiler that is easily adaptable to different SNAX cluster configurations. It simplifies key tasks such as memory allocation, synchronization, scheduling, and data management, enabling efficient multi-accelerator programmability.
\end{itemize}

The SNAX platform delivers a comprehensive solution for seamlessly integrating diverse hardware accelerators, fostering a flexible environment for developing and testing various multi-accelerator combinations in conjunction with software development. SNAX is available at \url{https://github.com/kuleuven-micas/snax_cluster}.

%% file: 2-background.tex

\section{Background and Motivation}
\label{sec:background}


\subsubsection*{A. Hardware For Heterogeneous Compute Systems}

In heterogeneous systems, several accelerator integration strategies exist that expose different trade-offs. There are two key interfaces in every accelerator: (1) \textit{a control interface}, for dispatching configurations, commands, and tasks to an accelerator; and (2) \textit{a data interface} for reading and writing data to memory. These interfaces can either tightly or loosely communicate with a CPU or memory \cite{esp_loose_vs_tight}.

\begin{figure}[t]
    \centering
    \includegraphics[width=1\columnwidth]{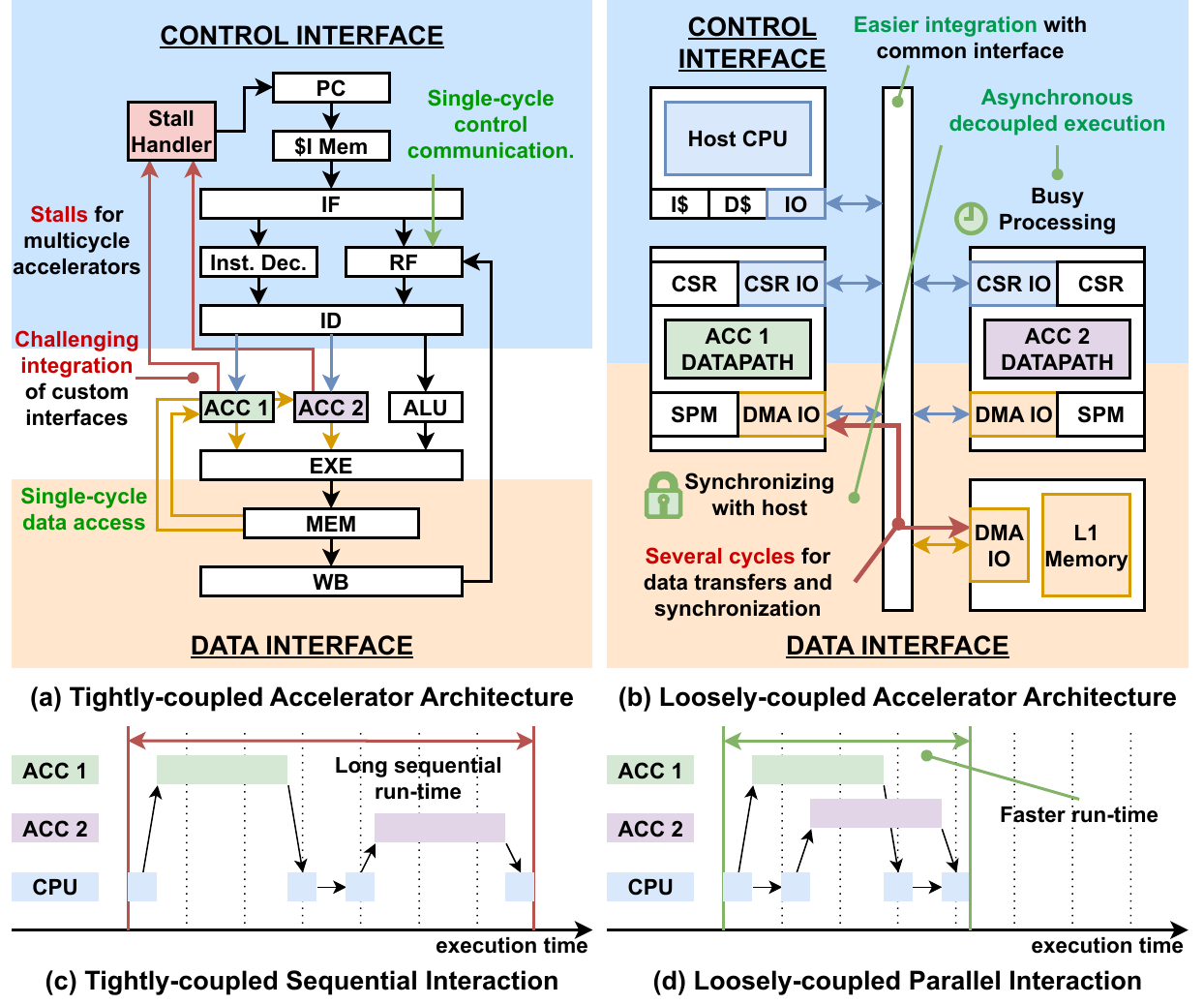}
    \caption{Comparison of tightly vs. loosely coupled architectures and execution times \cite{esp_loose_vs_tight}}
    \label{fig:tight_loose_background}
\end{figure}

A \textit{tightly coupled} accelerator \cite{acm_hetero_os_tight_coupling}, as shown in Figure \ref{fig:tight_loose_background}a, is integrated into the CPU pipeline as a coprocessor \cite{pulp_xpulpnn, pulp_snitch, pulp_spatz, chipyard_multi_hwacha}. The control interface uses an instruction decoder to offload instructions directly, while the data interface can directly access memory. The main advantage of this implementation is low-latency (single-cycle) access to both the CPU and memory.

However, pipeline stalls can occur when accelerators require multiple cycles for a task, leaving the CPU idle during execution (Figure \ref{fig:tight_loose_background}c) \cite{pulp_xpulpnn, chipyard_gemmini}. This limits parallel pipelined execution of accelerators, limiting overall efficiency.

Moreover, tightly coupled accelerators are challenging to integrate due to the required changes in the CPU pipeline and the need for interface wrappers. For example, these accelerators are hardcoded in the design and differ depending on the core processor to which they are attached to \cite{pulp_xpulpnn, ticsat_epfl}.

A \textit{loosely coupled} accelerator \cite{esp_loose_vs_tight}, as shown in Figure \ref{fig:tight_loose_background}b, connects its control and data interfaces to a common bus for communication with the CPU and memory. These accelerators interact indirectly through register- or memory-mapped interfaces \cite{pulp_xnor_hwpe, esp_12nm_agile, xheep_epfl, nvdla_accelerator}. This simplifies integration without needing to adapt to specific host system interfaces.

The key advantage of loosely coupled accelerators is their asynchronous decoupled execution \cite{isca_decoupled_access}, allowing the CPU and other accelerators to run tasks in parallel without stalling \cite{esp_agile_dev, nvidia_a100} (Figure \ref{fig:tight_loose_background}d). This asynchronous execution can achieve up to $30 \times$ better performance than tightly coupled accelerators where execution is mostly sequential \cite{esp_data_transfers}.

The main drawbacks in loosely coupled architectures are data transfer and synchronization overheads. Due to control mechanisms and limited DMA bandwidth (e.g., 64-bit per cycle in ESP \cite{esp_agile_dev}), these processes may require many cycles, causing memory conflicts and stalling of cores and accelerators. Furthermore, task scheduling and synchronization are challenging when accelerators have varying compute cycles: slower accelerators can bottleneck faster ones by delaying needed outputs, reducing overall efficiency\cite{overview_hpc_hardware_acc}. Overlapping DMA transfers, synchronization, and compute cycles can help alleviate these bottlenecks and improve system efficiency \cite{esp_agile_dev}.


\subsubsection*{B. Software For Heterogeneous Compute Systems}
The tight or loose coupling implementation strategies also affect software development. Tightly coupled architectures demand extensive compiler customization to support custom instructions for controlling the accelerator. For example, custom instructions manage tightly coupled accelerator units \cite{pulp_xpulpnn, ticsat_epfl}. These instructions are inserted manually through compiler directives in C code, often to build high-performance custom accelerator kernel libraries. This approach adds complexity for developers, especially for managing heterogeneous accelerators with different custom instructions. Moreover, without robust compiler support for these custom instructions, the accelerator remains inaccessible to non-hardware experts.

Loosely coupled architectures manually manage memory allocation, synchronization, scheduling, and data management \cite{esp_data_transfers, chipyard_gemmini}. If not done properly, overhead cycles due to stalls can reduce overall efficiency \cite{esp_agile_dev}; to handle this, some platforms use external scheduling platforms (e.g., RTOS) to help manage these tasks \cite{xheep_epfl}.


In summary, the lack of a cohesive hardware-software integration strategy limits the efficiency of a heterogeneous system. Tightly coupled accelerators offer low-latency interactions but at the cost of sequential execution and HW-SW compatibility issues. In contrast, loosely coupled accelerators enable asynchronous and parallel execution but suffer from data transfer and synchronization overheads if not properly managed. In general, system efficiency depends on the ease of integration and maximizing compute cycles over overhead cycles, while software manages data flows across various accelerators. A unified HW-SW framework is essential to streamline integration and ensure high performance.

%% file: 2.5-snax.tex
\section{SNAX: Hybrid Coupling}
\label{sec:hybrid_coupling}

\subsubsection*{A. The Hybrid-coupling Concept} To jointly exploit the benefits of the ease-of-integration of loosely-coupled accelerators, as well as the system level efficiency benefits of tightly coupled accelerators, SNAX proposes a new \textit{hybrid-coupling} approach: \textit{a loosely coupled control interface and a tightly coupled data interface}.
Figure \ref{fig:snax_hybrid_coupling} visualizes the conceptual framework of the hybrid coupling concept.

The loosely coupled control interface enables asynchronous parallel execution across multiple accelerators.
This approach allows a management system to offload tasks efficiently without stalling, enabling parallel accelerator execution. On the hardware side, this is achieved through a uniform control interface for configuring accelerators. Consequently, on the software side, this decouples control information about the accelerators allowing control tasks to be uniformly programmed simplifying configuration and management.

Conversely, the tightly coupled data interface ensures low-latency access to shared memory while catering to the unique data access patterns and bandwidth requirements of each accelerator. In hardware, this involves providing customized data interfaces that enable single-cycle memory access at different bandwidths. From a software perspective, decoupling the data flow information enables the integration of tailored data access patterns of an accelerator into the system's data management strategy, ensuring optimal performance.

\subsubsection*{B. System Level Parallel Operation}

\begin{figure}[t]
    \centering
    \includegraphics[width=0.9\columnwidth]{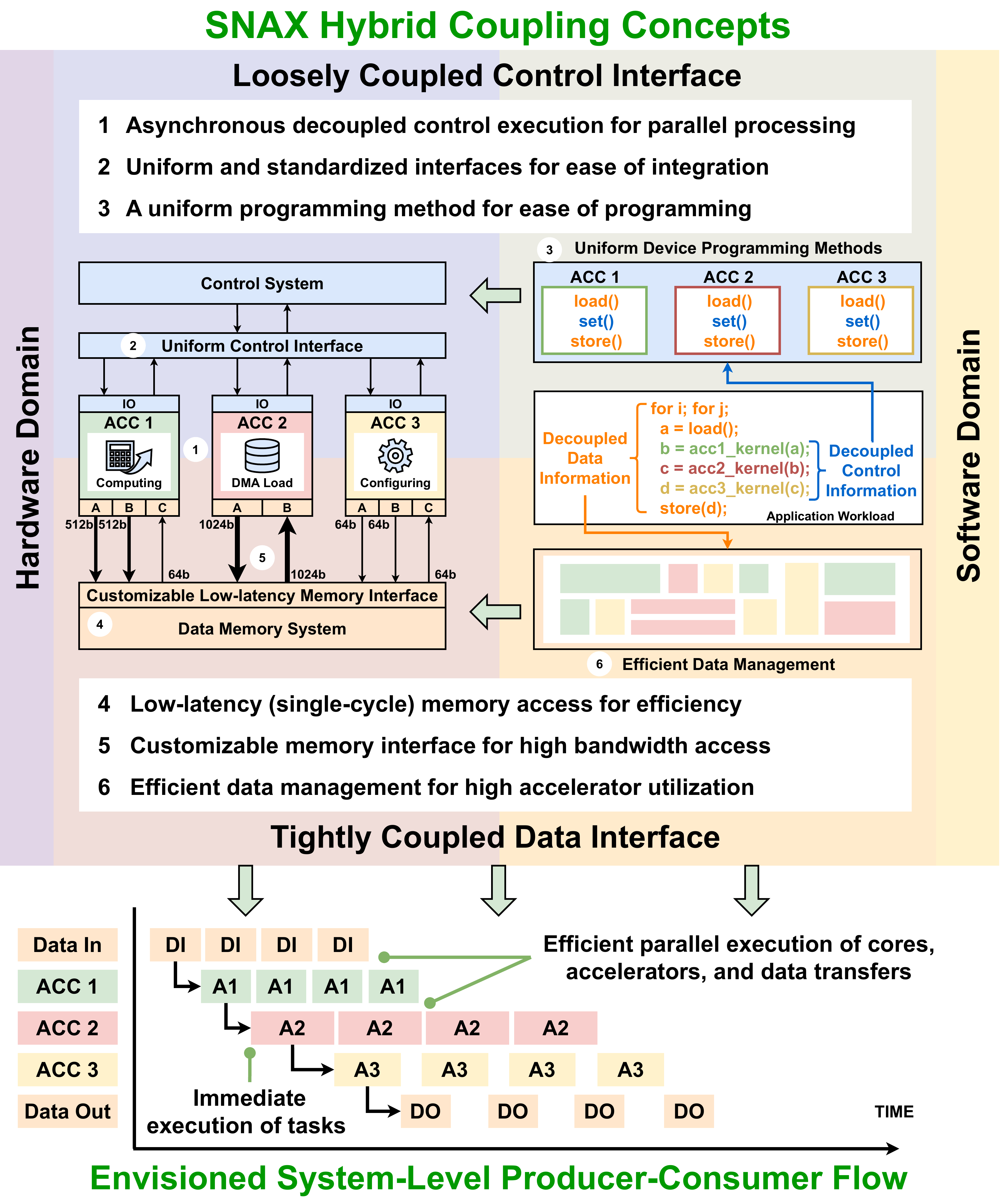}
    \caption{SNAX hybrid coupling concepts affecting both HW and SW development, and the envisioned system-level execution.}
    \label{fig:snax_hybrid_coupling}
\end{figure}

The hybrid-coupling concept enables multiple accelerators to operate in parallel within a shared memory system achieving high compute utilization and efficient data access. It facilitates a streamlined consumer-producer data flow across parallel accelerators, where data is seamlessly passed from one accelerator to the next without increasing software complexity for the designer, as illustrated in Figure \ref{fig:snax_hybrid_coupling}.

To realize these characteristics, this work presents the SNAX multi-accelerator compute cluster template (HW - Section \ref{sec:snax_cluster_arch}) and the complementary SNAX MLIR compile toolchain (SW - Section \ref{sec:snax_mlir}).

%% file: 3-hardware.tex
\section{SNAX-Cluster HW Architecture}
\label{sec:snax_cluster_arch}

\begin{figure}[t]
    \centering
    \includegraphics[width=0.9\linewidth]{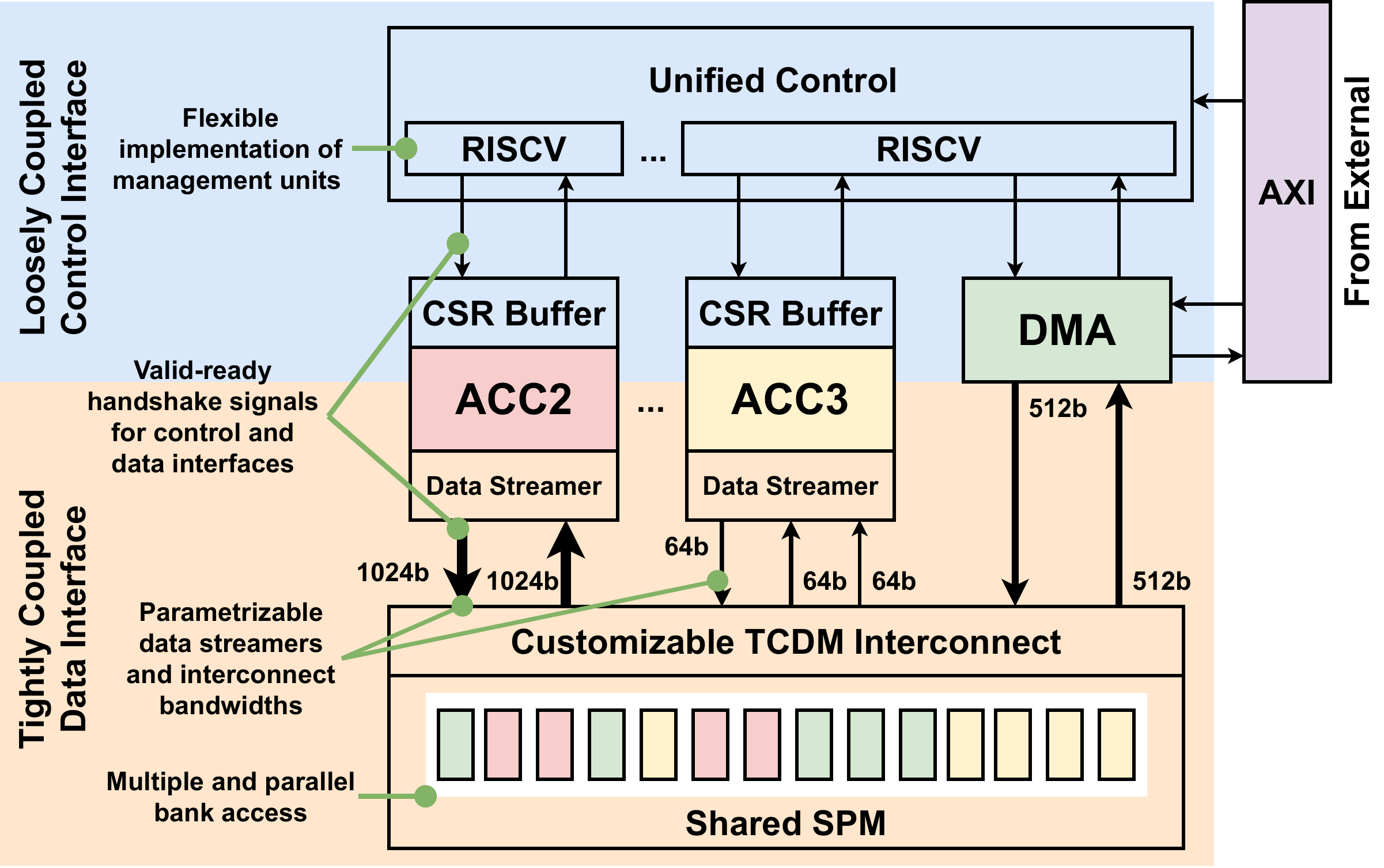}
    \caption{SNAX Multi-accelerator Compute Cluster Architecture}
    \label{fig:snax_architecture}
\end{figure}

Figure \ref{fig:snax_architecture} shows the SNAX multi-accelerator compute cluster's hardware architecture built on the hybrid-coupling framework's principles.
Our implementation of the loosely coupled control interface and the tightly coupled data interface aims to facilitate ease of integration while maximizing efficiency.

\subsubsection*{A. Implementing the Loosely Coupled Control Interface}

For the control interface, SNAX utilizes one or more lightweight, single-cycle RISC-V integer cores \cite{pulp_snitch} as management units. These cores are responsible for sending register-mapped control signals to the attached accelerators. Leveraging standard RISC-V Control and Status Register (CSR) instructions, the cores efficiently offload tasks to the accelerators in a "fire-and-forget" manner. Each core independently oversees one or more accelerators, enabling asynchronous, decoupled execution across the system.

The CSR interface between the RISC-V and accelerators consists of register write enable, address, and data ports synchronously managed by valid-ready signals \cite{axi_valid_ready}. This generic CSR interface makes it easy for any custom accelerator to comply with.

The adopted control approach standardizes low-level programming across accelerators: regardless of the accelerator type, configurations are set using uniform CSR read and write instructions while only register addresses vary, as each accelerator has unique addresses. This standardization greatly simplifies programming for heterogeneous accelerator systems (see Section \ref{sec:snax_mlir}).

The CSR interface includes double buffering to hide register setup time, allowing new configurations to be pre-loaded while accelerators execute their tasks. Since configuring registers can take multiple cycles depending on the number of parameters, the pre-loading hides setup latency, boosting system performance.

\subsubsection*{B. Implementing the Tightly Coupled Data Interface}

The SNAX architecture's tightly coupled data interface enhances data access efficiency through a configurable shared, multi-banked scratchpad memory (SPM) across all accelerators. This shared SPM enables single-cycle read and write operations with parallel access to multiple banks, reducing latency and enabling seamless data exchange between accelerators. Each accelerator connects via a customizable tightly coupled data-memory (TCDM) interconnect \cite{pulp_tcdm}. The bandwidth and the number of ports of the TCDM interconnect are adjustable at design time. The interconnect uses round-robin scheduling to handle bank contention, prioritizing higher-bandwidth ports. Each data port is also synchronized with valid-ready signals. The shared memory approach eliminates costly DMA transfers from accelerator to accelerator, enabling direct memory access for the accelerators across a shared address space.

To further optimize data flow, SNAX uses parametrizable data streamers at the accelerator-memory interface \cite{pulp_xnor_hwpe}.
These streamers have autonomous load/store address generation (configured via CSR) and FIFO buffers to manage memory conflicts, ensuring a smooth, continuous data stream into the accelerators at each cycle. To improve memory access efficiency, streamers include hardware loop support for generating target memory addresses towards optimized nested for-loop data access patterns \cite{zigzag_kul}. Design-time customizations allow for adjustable streamer bandwidth, for-loop structures, and FIFO depths, while loop counters can be configured at run time, supporting adaptable and efficient data access.

A user who plans to use the cluster can primarily focus on developing their accelerator's data path while knowing they have access to a simplified control interface and efficient data streamers for continuous data feeding. They only need to comply with a simplified register interface on the control side as well as a simplified data interface on the data side.

\subsubsection*{C. Peripherals}

SNAX uses an AXI network to transfer data from external sources into the SPM, with a high-bandwidth (512-bit) DMA for rapid data exchange. The implemented programmable DMA has two configurable strides  -- for source and destination -- and allows the management of 2D data transfers.

Synchronization across all cores, accelerators, and the DMA is ensured by a hardware barrier, which facilitates coordination between data transfers and accelerator tasks. These barriers are simple register fences that are set using CSR instructions.

The SNAX cluster architecture enables a continuous producer-consumer data flow, potentially achieving peak efficiency for a multi-core, heterogeneous, multi-accelerator system as envisioned in Figure \ref{fig:snax_hybrid_coupling}. By decoupling control and data interfaces, the proposed HW-SW development framework provides the compiler back-end with accelerators' components, such as the accelerator programmability (e.g. the supported kernel types) and the unique data access patterns available (e.g. optimized address generation for for-loops). 
By leveraging this information, the compiler can now efficiently map kernels and manage data flow, optimizing throughput and utilization across all accelerators.

%% file: 4-software.tex
\section{SNAX-MLIR SW infrastructure}
\label{sec:snax_mlir}

Operating multiple accelerators in a pipelined, streaming fashion would enable high hardware utilization but managing this flow poses significant challenges due to the diverse accelerators' programmabilities and data access patterns.
To address this, SNAX incorporates an MLIR-based compiler toolchain, which is customizable to various SNAX Cluster instantiations and accelerator combinations.
SNAX-MLIR automates tasks like managing reusable accelerator components, streaming data transfers, scratchpad management, and asynchronous execution, ensuring efficient operation across all accelerators in the system.
The deterministic nature of embedded AI applications, combined with a manually managed memory system and tightly integrated accelerator interfaces, enables the compiler to perform highly efficient, fine-grained optimizations on the source code.

\begin{figure}[t] 
    \centering
    \includegraphics[width=0.9\linewidth]{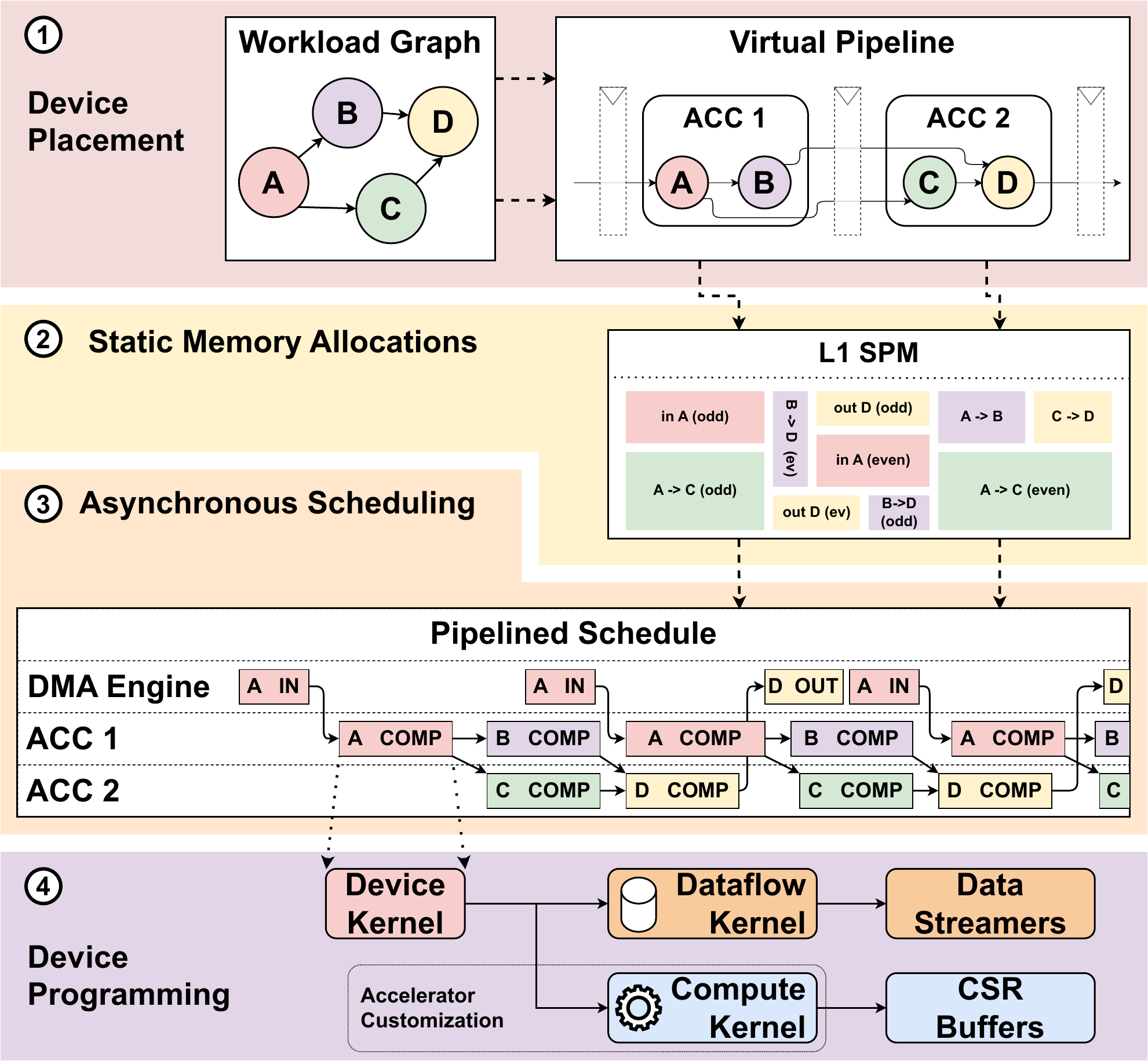} 
    \caption{Automated Pipelining in SNAX-MLIR.}
    \label{fig:wide_figure}
\end{figure}

This section dives into four key concepts of the SNAX-MLIR compiler, visualized in Figure \ref{fig:wide_figure}. Each concept contributes to the same overarching goal: enabling highly efficient code generation while simplifying the complexities of heterogeneous hardware-software integration. These concepts are demonstrated here in the context of achieving a pipelined consumer-producer flow, although they can be applied to various other execution scenarios as well.

\subsubsection*{Device Placement}

SNAX-MLIR offloads computation sections to the most suited accelerator based on workload characteristics. Each workload is decomposed into sub-computations, which are then assigned to accelerators based on their control and kernel descriptions. For workload sections that are incompatible with the available accelerators, the accompanying RISC-V core handles execution, minimizing off-cluster data movement. The asynchronous control model results in a scheduled virtual pipeline (Figure \ref{fig:wide_figure}.1), enabling parallel execution towards efficient hardware utilization.

\subsubsection*{Static Memory Allocation}

SNAX-MLIR allocates buffers in shared memory to support producer-consumer relationships without the need for intermediate memory transfers. This is achieved based on device placements determined in the previous stage and the data dependencies within the workload graph. Double buffering in the SPM enables pipelined execution, with separate buffers designated for reading and writing during alternating odd and even pipeline cycles.

\subsubsection*{Asynchronous Scheduling}

While SNAX's asynchronous control model offers high efficiency, managing it manually at a fine-grained level can be challenging. SNAX-MLIR simplifies this process by unrolling the virtual pipeline stages and inserting synchronization barriers between stages with data dependencies. This approach ensures tasks are executed in the correct order while enabling concurrency wherever possible. The system supports pipelined accelerator execution and allows overlapping DMA transfers with computation, maximizing accelerator utilization, especially in memory-bound workloads.

\subsubsection*{Device Programming}

Finally, the compiler generates accelerator-specific kernels. This is realized by producing CSR-read and CSR-write instructions that program all RISC-V hosts connected to one or more accelerators. Using the CSR-based approach ensures a standardized and uniform method for managing an accelerator's computation and data flow tasks.

Each kernel is divided into two components —compute and dataflow kernels— aligned with SNAX's hybrid-coupling strategy. The compute kernel contains unique CSR configurations to define the accelerator's functionality and execution tasks. Meanwhile, the dataflow kernel is generated based on planned static memory allocations and the accelerator's access patterns, programmed into the accelerator's data streamers. These kernels enable SNAX-MLIR to create pipelined, multi-accelerator schedules adaptable to any combination of accelerators, ensuring efficient and coordinated execution. The tasks or workloads offloaded to the accelerators are fixed at compile time.

%% file: 5-evaluation.tex
\section{Evaluation}

\subsubsection*{A. Experimental Setup}

\begin{figure}[t]
    \centering
    \includegraphics[width=0.9\columnwidth]{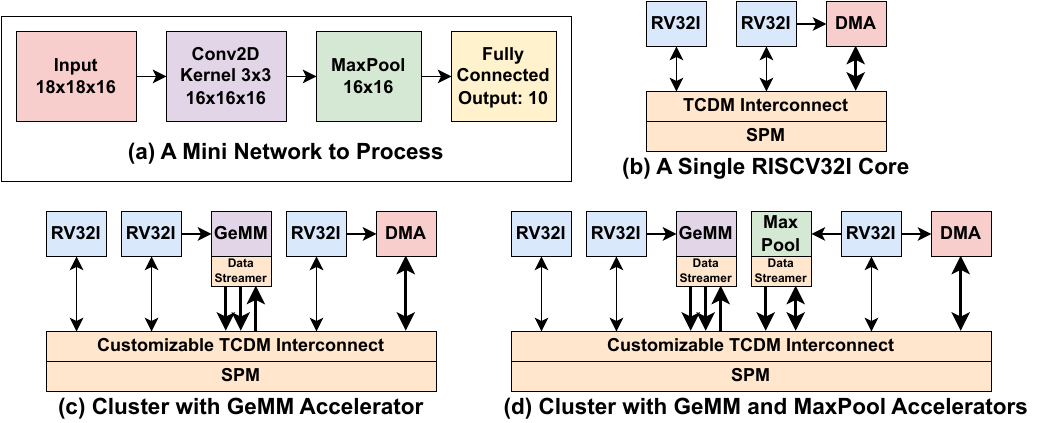}
    \caption{Workload to process and the different SNAX cluster architectures}
    \label{fig:eval_workload_archs}
\end{figure}

To demonstrate the integration flexibility and performance of the SNAX platform, we configure SNAX with various accelerator combinations and deploy several machine learning workloads. The first experiment evaluates the platform's overall efficiency by examining area, latency, and power. This evaluation uses a simplified artificial workload with representative machine learning layers—a convolutional layer, a max-pooling layer, and a fully connected layer—all operating at 8-bit precision, as shown in Figure \ref{fig:eval_workload_archs}a. Afterward, we deploy tiled matrix multiplication to asses as well as end-to-end workloads from the MLPerf Tiny v1.0 benchmark to demonstrate the platform’s utilization and ease of programmability.

For all experiments, performance metrics are obtained through cycle-accurate RTL simulations conducted with Verilator and Questasim. The architectures are synthesized at 800 MHz using TSMC 16nm technology and evaluated for area using Synopsys DC Compiler. Power metrics are measured from switching-annotated post-synthesis netlist simulations, analyzed with Synopsys PrimeTime.

Figures \ref{fig:eval_workload_archs}b, \ref{fig:eval_workload_archs}c, and \ref{fig:eval_workload_archs}d illustrate the different hardware platform configurations used in our experiments, with a varying number of accelerators.

\subsubsection*{B. Flexible Heterogeneous Integration}

We first assess the effort and cost of integrating an accelerator into the system. Key design-time customizations for the cluster include attaching new accelerators to RISC-V cores, adjusting TCDM interconnect parameters, and configuring data streamers. We begin with a baseline SNAX cluster architecture containing a single RISC-V32I core capable of running the entire workload, shown in Figure \ref{fig:eval_workload_archs}b.

To improve the performance of convolutions, the second SNAX cluster introduces a GeMM accelerator \cite{open_gemm_kul_2024} optimized for CNN kernels, as shown in Figure \ref{fig:eval_workload_archs}c. This accelerator includes 512 processing elements (PEs) and can process $8 \times 8 \times 8$ matrices in a single cycle, with 512-bit streaming bandwidth for both input matrices (A and B) and a 512-bit output streaming bandwidth. Figure \ref{fig:eval_workload_archs}d further expands the architecture by adding a max-pool accelerator to accelerate the max-pooling kernel, supporting 8 parallel max-pool kernels with configurable kernel size and 512-bit input/output streaming bandwidth.

\begin{figure}[t]
    \centering
    \includegraphics[width=\linewidth]{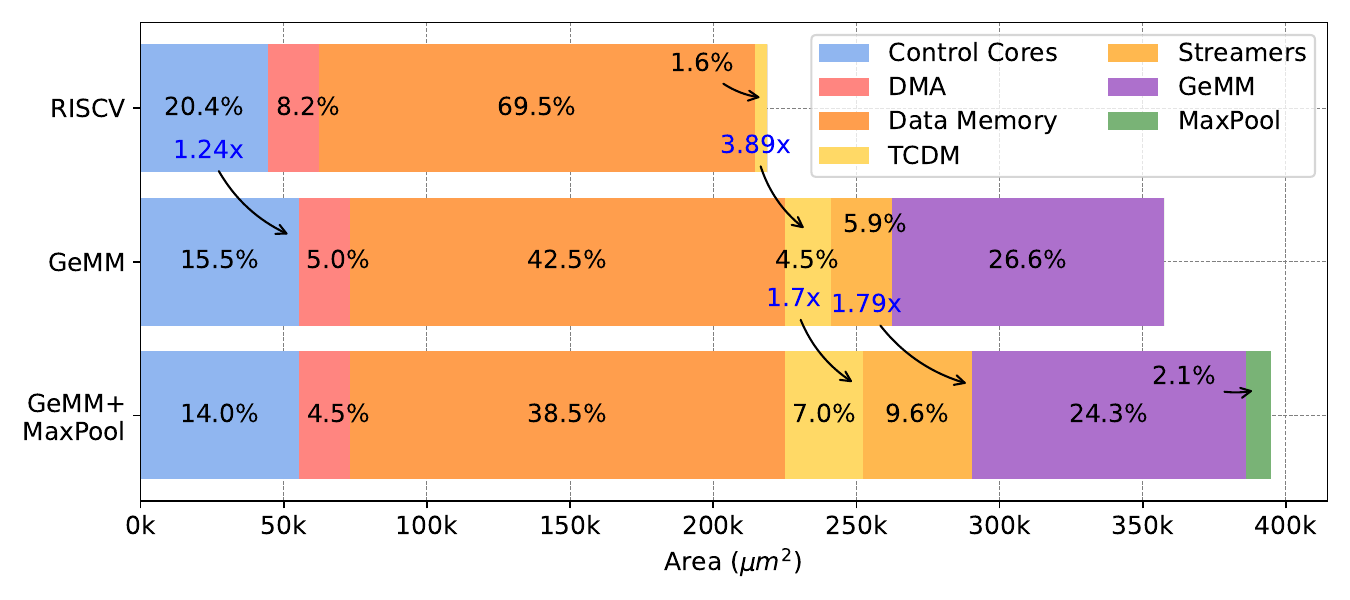}
    \caption{Area breakdown of each architecture}
    \label{fig:area_architecture}
\end{figure}

Figure \ref{fig:area_architecture} shows the area breakdown for each architecture. The control cores area includes the RISC-V cores and instruction memory. Adding a RISC-V core increases the control area by 1.17× (from \ref{fig:eval_workload_archs}b to \ref{fig:eval_workload_archs}c), while sharing an accelerator with an existing core has minimal impact on the area (from \ref{fig:eval_workload_archs}c to \ref{fig:eval_workload_archs}d). On the data interface side, the TCDM interconnect area expands to accommodate the high bandwidth demands of the accelerators (i.e., the GeMM adds additional 2 512-bit read ports and one 2,048-bit write port, and the maxpool only adds 2 512-bit ports). Additionally, data streamers introduce a notable area increase, supporting the accelerators' specific data access patterns and buffering.

All customizations within the platform are managed through a single configuration file, with parameters for control and data interfaces. For the control interfaces, accelerators can be either dedicated to or shared among one or more RISC-V cores. For instance, \ref{fig:eval_workload_archs}c illustrates a setup where the GeMM accelerator is attached to a single RISC-V core, while \ref{fig:eval_workload_archs}d demonstrates a configuration where the same core is shared to control both the Max-pool and DMA accelerators. On the data interfaces, the data streamers and TCDM interconnect can be configured to meet the specific bandwidth requirements of each accelerator. The architecture is generated using templates and hardware generators, with build times taking just a few minutes (approx. 3 mins) with open-source simulators such as Verilator.


\subsubsection*{C. Efficient Heterogeneous Acceleration}

To demonstrate the efficiency of SNAX, we run the network from Figure \ref{fig:eval_workload_archs}a on the architecture in Figure \ref{fig:eval_workload_archs}d to evaluate how the addition of each accelerator influences the system throughput.

\begin{figure}[t]
    \centering
    \includegraphics[width=0.9\linewidth]{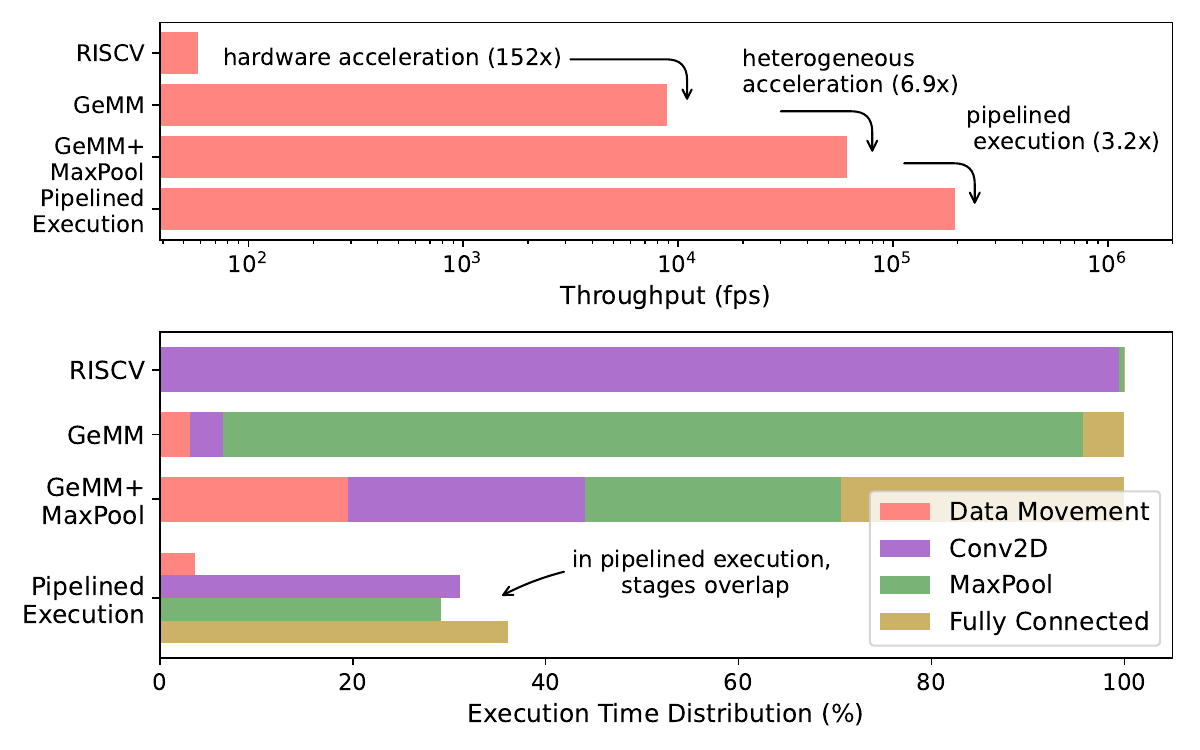}
    \caption{SNAX Performance for Heterogeneous acceleration.}
    \label{fig:evaluation-toy}
  \end{figure}

As a baseline, the entire network is deployed on the RISC-V core, which provides performance comparable to other low-power, single-issue microcontrollers. Figure \ref{fig:evaluation-toy} shows the network’s average throughput and the cycle distribution for each layer. The convolution layer takes majority of the execution time.

To accelerate this, we add the GeMM accelerator in the system and specify how to program it for dispatching convolution layers, achieving a $152 \times$ performance boost. However, as Figure \ref{fig:evaluation-toy} shows, the system now has bottlenecks in the max-pooling layer. Activating the max-pooling accelerator for this workload yields an additional $6.9 \times$ throughput improvement. At this point, the execution times of all layers are well-balanced. Despite this balance, accelerators remain underutilized due to the layer-by-layer sequential execution of the model involving data transfers between layers. 

Combining the loose asynchronous control model with a tightly coupled shared memory system enables efficient producer-consumer pipelines to overcome this underutilization. This setup parallelizes the convolutional layer by the GeMM accelerator, the max-pooling layer by the max-pooling accelerator, the fully connected layer by a RISC-V core, and data movement by the DMA engine, and yields an additional $3.18 \times$ throughput increase. Notably, in all these scenarios, the source code of the original program remains unchanged. Only the new targets need programming instructions to execute their specific layers. The automated compiler passes manage dispatching, synchronization, data movement, and pipelined execution seamlessly. 
The compiler determines whether to enable pipelined execution or default to sequential execution based on explicit configuration flags and target descriptions provided during compilation, ensuring the correct execution model is applied for the given setup.

\begin{figure}[t]
    \centering
    \includegraphics[width=0.95\columnwidth]{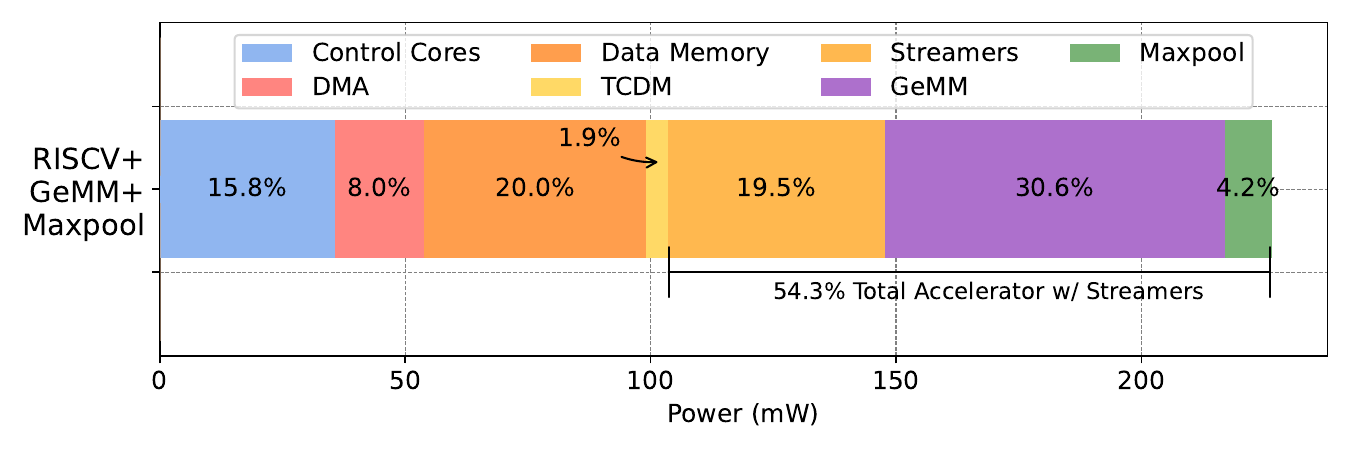}
    \caption{Power breakdown for parallel processing}
    \label{fig:power_sample}
\end{figure}

Figure \ref{fig:power_sample} shows the power distribution per component during parallel execution. The majority of power consumption is consumed by the accelerators and their streamers, followed by data memory access, peripheral interconnect, and RISC-V cores.

\subsubsection*{D. Chasing Peak Utilization}

Maintaining high accelerator utilization is a challenge for accelerator system integration. In several cases, accelerators exhibit excellent peak performances, yet, when integrated into an SoC they fail to perform close to this peak for real workloads. This loss is primarily due to data transfers and control synchronization.
The next experiment will demonstrate how SNAX pushes both the accelerator and the system to their limits, across a wide range of workload arithmetic intensities, evaluated through a roofline model \cite{roofline} of the architecture. For this, we benchmark the system of Figure 6(b) with a variety of tiled matrix multiplications. For each tile, input data is transferred into the system via the 512-bit AXI bus, processed by the accelerator, and the partial result is sent back. By sweeping the tile sizes, the arithmetic intensity (operations per byte) of the workload changes allowing us to evaluate performance against the roofline model of the system.
As a comparison, we implement the same experiment using the available C runtime libraries \cite{open_gemm_kul_2024} using the same GeMM accelerator.
This implementation, relying solely on conventional integration methods and requiring a similar programming effort, serves as a baseline to demonstrate the performance achievable without leveraging the hybrid integration capabilities of SNAX.

In high arithmetic intensity scenarios (large tile sizes), where accelerator performance is the bottleneck, SNAX achieves 92\% PE utilization. For low arithmetic intensity, limited by AXI bandwidth (data transfer bandwidth), SNAX maintains an average utilization of 79\% of the available AXI bandwidth. At the ridge point, where the workload balance between data movement and computation is equal, the asynchronous control model and pipelined execution in SNAX still enable 78\% utilization. This clearly shows that even in scenarios involving single-kernel execution, the hybrid coupling in SNAX proves advantageous, highlighting the system's ability to effectively maximize resource utilization across workloads with widely diverging computational and data-intensity characteristics.

\begin{figure}[t]
    \centering
    \includegraphics[width=0.90\linewidth]{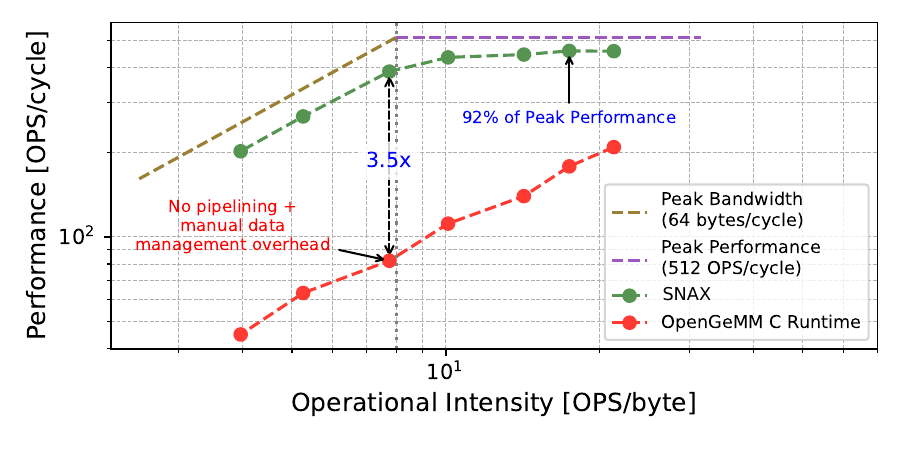}
    \caption{Roofline model \cite{roofline} for SNAX.}
    \label{fig:roofline}
\end{figure}

\subsubsection*{E. End-to-end execution}

\begin{table}[t]
    \caption{SotA Comparison with Other Heterogeneous Architectures}
    \label{tbl:sota_compare}
    \resizebox{\columnwidth}{!}{ 
    \begin{tabular}{|c|c|c|c|c|c|c|c|}
        \hline
        \rowcolor[HTML]{C0C0C0} 
        \textbf{System} & \begin{tabular}[c]{@{}c@{}} \textbf{ESP} \\ \cite{esp_isscc_2024} \end{tabular} &  \begin{tabular}[c]{@{}c@{}} \textbf{ChipY.} \\ \cite{chipyard_multi_core_multi_acc} \end{tabular} &  \begin{tabular}[c]{@{}c@{}} \textbf{Sira.} \\ \cite{pulp_siracusa} \end{tabular} \tnote{*} & \begin{tabular}[c]{@{}c@{}} \textbf{ST} \\ \cite{st_micro_controller, tiny_ml} \end{tabular}  & \begin{tabular}[c]{@{}c@{}} \textbf{GAP9} \\ \cite{gap_9_cluster, tiny_ml} \end{tabular} &  \begin{tabular}[c]{@{}c@{}} \textbf{DIANA} \\ \cite{diana_kul_2023, htvm_kul_2023} \end{tabular} & \textbf{SNAX*} \\ \hline
        
        \textbf{Technology (nm)} & 12 & 22 & 16 & 40 & 22 & 22 & 16 \\ \hline
        
        \textbf{Frequency (MHz)} & 1,600 & 960 & 360 & 120 & 260 & 340 & 800 \\ \hline
        
        \textbf{Memory (B)} & 7.7M & 1M & \begin{tabular}[c]{@{}c@{}} 256k (L1)\\ 6M (L2) \end{tabular} & 2.6M & 3.6M &\begin{tabular}[c]{@{}c@{}} 320k (L1) \\ 512k (L2)\end{tabular} & 128k \\ \hline
        
        \textbf{Area (mm\textsuperscript{2})} & 64 & 16 & 16 & - & - &8.91 & 0.45 \\ \hline
        
        \begin{tabular}[c]{@{}c@{}} \textbf{Area Efficiency} \\\textbf{(TOP/s/mm\textsuperscript{2})} \end{tabular} & - & - & 0.065 & - &  -& \begin{tabular}[c]{@{}c@{}}12.88 (A) **\\ 3.33 (AD) \end{tabular} & 0.924 \\ \hline
        
        \begin{tabular}[c]{@{}c@{}} \textbf{Total Power} \\\textbf{(mW)} \end{tabular} & 4,300 & - & 332 & 29.7 & 50 & \begin{tabular}[c]{@{}c@{}} 31 (D) \\ 54 (A) \end{tabular} & 227 \\ \hline \hline
        
        \begin{tabular}[c]{@{}c@{}} \textbf{Latency (ms)} \\\textbf{ToyAdmos} \end{tabular} & - & - & - & 7.75 & 0.18 & 0.36*** & \color{blue}{0.024} \\ \hline
        
        \begin{tabular}[c]{@{}c@{}} \textbf{Latency (ms)} \\\textbf{ResNet-8} \end{tabular} & - & - & - & 227 & 0.62 & 1.19*** & \color{blue}{0.132} \\ \hline
        
        \begin{tabular}[c]{@{}c@{}} \textbf{Energy (uJ)} \\\textbf{ToyAdmos} \end{tabular} & - & - & - & 230 & 9 & \begin{tabular}[c]{@{}c@{}} 11 (D) \\ 19 (A) \end{tabular} & \color{blue}{5.16} \\ \hline
        
        \begin{tabular}[c]{@{}c@{}} \textbf{Energy (uJ)} \\\textbf{ResNet-8} \end{tabular} & - & - & - & 6.7k & 31 & \begin{tabular}[c]{@{}c@{}} 37 (D) \\ 64 (A) \end{tabular} & \color{blue}{28} \\ \hline \hline
        
        \begin{tabular}[c]{@{}c@{}} \textbf{Control} \\\textbf{Coupling} \end{tabular} & Loose & Loose & Tight & Tight & Tight & Loose & Loose \\ \hline
        
        \begin{tabular}[c]{@{}c@{}} \textbf{Data} \\\textbf{Coupling} \end{tabular} & Loose & Loose & Tight & Tight & Tight & Loose & Tight \\ \hline

        \begin{tabular}[c]{@{}c@{}} \textbf{Accelerator} \\ \textbf{Streamer} \\ \textbf{Support} \end{tabular} & \color{red}{\ding{55}} & \color{red}{\ding{55}} & \color{darkgreen}{\ding{51}} & \color{red}{\ding{55}} & \color{red}{\ding{55}} & \color{red}{\ding{55}} & \color{darkgreen}{\ding{51}} \\ \hline
        
        \begin{tabular}[c]{@{}c@{}} \textbf{Compiler} \\ \textbf{Support} \end{tabular} & \color{red}{\ding{55}} & \color{darkgreen}{\ding{51}} & \color{red}{\ding{55}} & \color{red}{\ding{55}} & \color{darkgreen}{\ding{51}} & \color{darkgreen}{\ding{51}} & \color{darkgreen}{\ding{51}} \\ \hline
        
        \begin{tabular}[c]{@{}c@{}} \textbf{HW-SW} \\ \textbf{Co-Development} \end{tabular} & \color{red}{\ding{55}} & \color{red}{\ding{55}} & \color{red}{\ding{55}} & \color{red}{\ding{55}} & \color{red}{\ding{55}} & \color{red}{\ding{55}} & \color{darkgreen}{\ding{51}} \\ \hline
    \end{tabular}
    }
    
    \begin{tablenotes}
        \begin{scriptsize}
            \item[*] * SNAX results are synthesized results only and exclude IO area and power.
            \item[**] ** For DIANA, A pertains to the analog IMC, and D pertains to the digital component.
            \item[***] *** Measurements were taken from \cite{htvm_kul_2023}
        \end{scriptsize}
    \end{tablenotes}
\end{table}

With SNAX's MLIR-based compiler, machine learning frontends like TensorFlow and PyTorch can directly deploy neural networks onto SNAX platforms. In the context of low-power computing, we deployed a Deep Autoencoder from the MLPerf Tiny v1.0 Benchmarking suite using TensorFlow Lite's MLIR Importer on the system of Figure \ref{fig:eval_workload_archs}d. Table \ref{tbl:sota_compare} shows the performance compared to other low-power microcontrollers and AI accelerators with available data.

SNAX demonstrates significant gains, even over AI-accelerated devices like GAP9 and DIANA, with $7.5 \times$ and $15 \times$ speedups respectively. This suggests possible underutilization or limitations in these platforms compared to SNAX’s hybridly coupled architecture.

\subsubsection*{F. Comparison with SotA}

Finally, Table \ref{tbl:sota_compare} compares the SNAX architecture in Figure \ref{fig:eval_workload_archs}d to other state-of-the-art multi-accelerator architectures and platforms. Despite its limited 128 kB memory, simplified RISC-V cores, and basic accelerators, SNAX performs strongly on TinyML workloads. Its high frequency, small area, high performance, and energy efficiency surpass other platforms. This efficiency stems from optimized data access, asynchronous parallel execution of accelerators, and compiler-managed data layout and scheduling. The results demonstrate that a co-developed hardware-software framework can achieve excellent overall system efficiency.

%% file: 6-conclusion.tex
\section{Conclusion}

SNAX provides a flexible hardware-software framework for integrating diverse accelerators in multi-caccelerator clusters. Its hybrid coupling enables uniform control and efficient, customizable data management. Combined with the SNAX-MLIR compiler, it maximizes performance and utilization. Experimental results on a low-power heterogeneous SoC show over $10 \times$ improvements in latency and energy efficiency with more than 90 \% utilization achieved through compiler-driven data access, layout and scheduling.